\documentclass[epj]{svjour}
\usepackage{graphics}
\usepackage{epstopdf}
\usepackage{amsmath}
\begin{document}
\title{Gravity effects on mixing with magnetic micro-convection in microfluidics\thanks{Contribution to the Topical Issue “Flowing Matter, Problems and Applications” edited by Federico Toschi, Ignacio Pagonabarraga Mora, Muno Araujo, Marcello Sega.}}

\author{G. Kitenbergs\inst{1}$^,$\thanks{\email{guntars.kitenbergs@lu.lv}} \and A. Tatu\c l\v cenkovs\inst{1}$^,$\thanks{\email{andrejs.tatulcenkovs@lu.lv}} \and L. Pu\c kina\inst{1} \and A. C\=ebers\inst{1,2}
}                     
%
%
\institute{MMML lab, Department of Physcis, University of Latvia, Riga, LV-1002, Latvia, \and Chair of Theoretical Physics, Department of Physics, University of Latvia, Riga, LV-1002, Latvia}
\date{Received: date / Revised version: date}
%
\abstract{
Mixing remains an important problem for development of successful microfluidic and lab-on-a-chip devices, where simple and predictable systems are particularly interesting. One is magnetic micro-convection, an instability happening on the interface of miscible magnetic and non-magnetic fluids in a Hele-Shaw cell under applied field. Previous work proved that Brinkman model quantitatively explains the experiments. However, a gravity caused convective motion complicated the tests. Here we first improve the experimental system to exclude the parasitic convection. Afterwards, we experimentally observe the magnetic micro-convection, finding and quantify how gravity and laminar flow stabilizes the perturbations that create it. Accordingly, we improve our theoretical model for a zero-flow condition and perform linear analysis. Two dimensionless quantities - magnetic and gravitational Rayleigh numbers - are used to compare the experimental observations and theoretical predictions for the critical field of instability and the characteristic size of the emerging pattern. Finally, we discuss the conditions at which gravity plays an important role in microfluidic systems.
} 
\maketitle
\section{Introduction}
\label{intro}
Microfluidics and lab-on-a-chip devices \cite{RevMicFluid} have been an active research topic for the last years.
Although many microfluidics fabrication routines and possible applications have been demonstrated, one of the key challenges remains the same - the physical limitations for mixing.
As microfluidics deals with manipulation of liquids in narrow channels, small Reynolds numbers and laminar flows are typical, where mixing happens only due to the slow diffusion process.
To speed up the process, passive and active micromixers can be used \cite{RevMicMix}.
A part of active mixers are based on magnetic materials and fields, which has opened a sub-field called micro-magnetofluidics \cite{RevMicMagFluid}.
Many different ways for mixing with magnetic elements have been proposed \cite{RevMagMix}.
One of them is to use magnetic micro-convection \cite{JMMM}.
In contrary to many others, this method has been studied extensively from the physical point of view and has a well developed theoretical model \cite{MHD,JFM1,JFM2}.
Here we extend this study by considering gravity effects.

The magnetic micro-convection is caused by a ponderomotive force, which acts on the magnetic fluid in a homogeneous applied field.
The ponderomotive force is proportional to two factors.
The concentration of the magnetic particles and the local gradient of the magnetic field, which arises from the self-magnetic field of the magnetic liquid.
Moreover, the force is potential only when the concentration gradient is collinear to the magnetic field gradient.
A flow is therefore created by any concentration perturbation that destroys this collinearity.
It is important to note, that this happens only if the magnetic field is higher than a critical value.

Several characteristics of the magnetic micro-convection were found by a linear stability analysis, also showing the importance of the initial smearing of the interface~\cite{Igonin}.
Also basic experimental characterization of the critical field and characteristic wavelength has been done~\cite{Derec}.
The instability has been also studied on a circular interface \cite{Kuan} and recently extended to characterizing secondary waves \cite{Wen} and a rotating system \cite{Wen2}.

In comparison, in our previous work \cite{JFM2} we showed a quantitative agreement between the experiments and the Brinkman model of the magnetic micro-convection.
However, we had to introduce an effective diffusion coefficient to take into account the extraordinary quick smearing of the interface without any magnetic field.
The experiment used a microfluidics cell that was placed horizontally, with magnetic field and gravity pointing perpendicular to the cell.
It was later proved that the smearing was actually a convective motion within the cell, that arises from the small density difference between the magnetic and non-magnetic fluids. 
This difference forces the slightly denser magnetic fluid to flow under the non-magnetic, what, when viewed in a 2D microscope from above, resembles diffusive process~\cite{GKthesis}.
Here we eliminate this convective flow by turning the system sideways and putting the slightly denser fluid below.

The following paper is organized in the following way.
Sect.~\ref{sec:2} introduces the modified experimental setup, describes a verification measurement with magnetic nanoparticle diffusion, as well as provides the information on experimental measurements that have lead to the understanding and characterization of the gravitational effects in magnetic micro-convection.
It also includes experimental measurements of the critical field and characteristic wavelength, which are approximated for a zero-flow case.
In sect.~\ref{sec:3} the theoretical model is updated to take into account a gravitational component and do a linear stability analysis, which gives information on the conditions for instability formation and its characteristics.
The comparison of experimental and theoretical results, as well a discussion is done in sect.~\ref{sec:5}, followed by main conclusions in sect.~\ref{sec:6}.

\section{Experimental system and observations}
\label{sec:2}
\subsection{Experimental setup}

\begin{figure}
\resizebox{1\columnwidth}{!}{\includegraphics{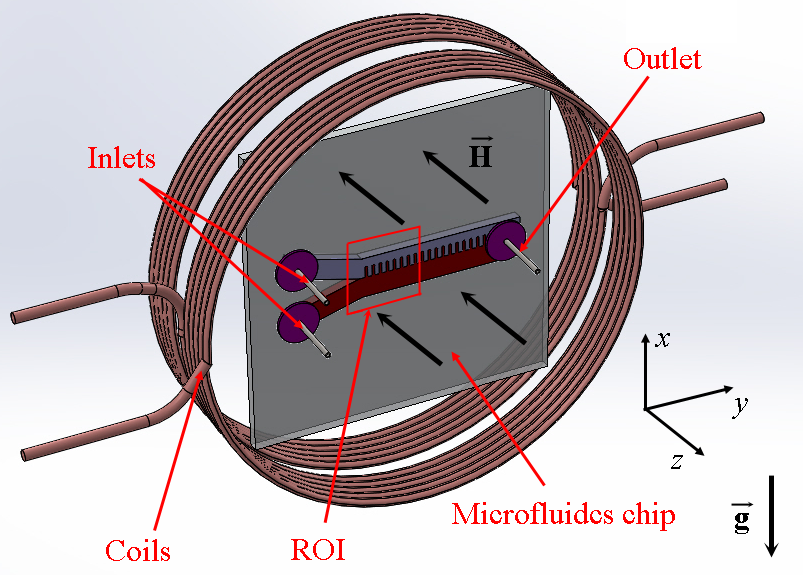}}
\caption{(color online) An illustration of the microfluidics chip within an electromagnet. The chip has a Y shaped channel with two inlets and one outlet. The lower inlet is for the denser magnetic fluid, while the upper inlet is for water. Coils provide a homogeneous magnetic field $H$, perpendicular to the chip. Region of interest (ROI) indicates the field of view of the camera. }
\label{fig:1}       
\end{figure}

The experimental setup mainly consists of four parts: an optical microscope (Zeis Stemi 2000-C) with a panel LED (Visional\textsuperscript \textregistered, $4$~W, $400$~lm, $3000$~K) as a light source, a syringe pump (Harvard Aparatus PHD Ultra) and a camera (Lumenera Lu165c, $15$~Hz) connected to a computer and an electromagnet around the microfluidics chip itself.
The microscope is put horizontally, to observe the vertically placed microfluidics chip.
A 3D printed holder (Mass Portal Pharaoh XD 20) fixes the chip in place within the electromagnet, that is made from two identical coils (for a simple illustration, see fig.~\ref{fig:1}).

The microfluidics chip is made of two microscope glass slides and a Parafilm~M\textsuperscript{ \textregistered} spacer.
Three holes are drilled in the top glass slide.
Then a metal tube from a cut syringe tip is glued in each of these holes.
This provides connections for tubing - 2 inlets (from syringe pump) and 1 outlet.
A Y shape is cut in the Parafilm~M\textsuperscript{ \textregistered} spacer, which is $h=0.13$~mm thick, with a paper knife.
Afterwards the spacer is welded between the two glass slides on a hot plate (Biosan MSH-300) at $75^{\circ}$~C, maintaining the original thickness.

In experiments we use two fluids - water based magnetic fluid, as described further, and distilled water as a non-magnetic miscible fluid.
The original magnetic fluid is made by a co-precipitation method \cite{FF}, forming maghemite nanoparticles which are stabilized with citrate ions and have a volume fraction $\Phi=2.8\%$,
Nanoparticles have an average diameter $d=7.0$~nm, saturation magnetization $M_{sat}=8.4$~G and magnetic susceptibility $\chi_m=0.016$, as determined by a vibrating sample magnetometer (Lake Shore 7404).

\begin{figure}
\resizebox{1\columnwidth}{!}{\includegraphics{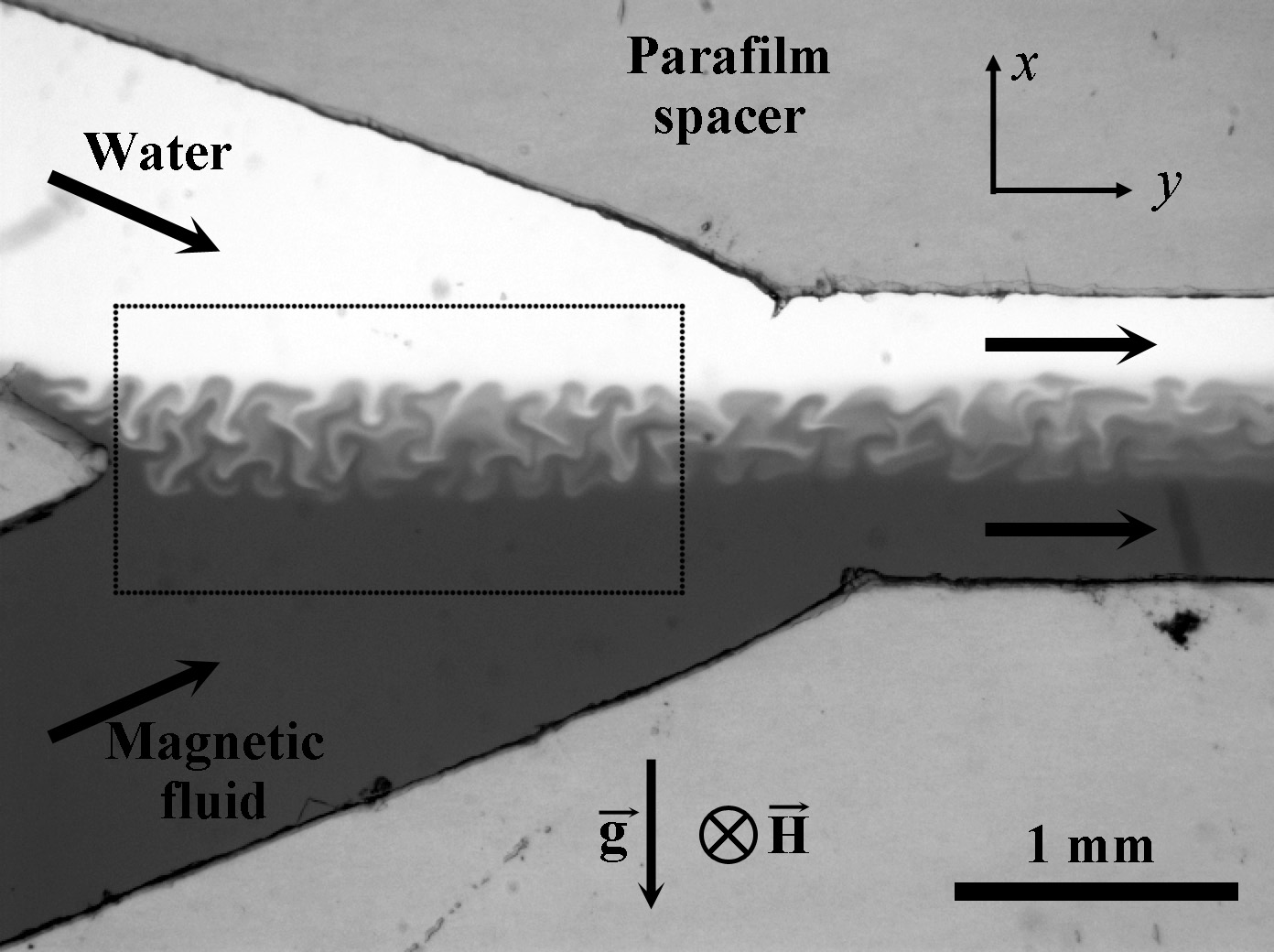}}
\caption{An image of the region of interest (ROI) of the microfluicis chip, where magnetic micro-convection is observed. Dotted rectangle is the area used in further analysis.}
\label{fig:2}       
\end{figure}

For liquid handling we use two $1$~ml syringes that are connected to the chip with FEP tubing (inner diameter $0.76$~mm, outer diameter $1.59$~mm, IDEX).
To keep the fluid interface stable in the microfluidics channel, shown in fig.~\ref{fig:2}, the denser magnetic fluid tubing is connected to the lower inlet, while water is connected to the upper inlet.
Once a sufficient magnetic field is applied, magnetic micro-convection emerges.
Coils are powered with a power supply (TENMA 72-2930) in a constant current mode and can create a homogeneous magnetic field up to $H=200$~Oe in the direction perpendicular to the plane of the chip.

To vary the density difference between the two fluids, we dilute the magnetic fluid with distilled water.
The density is calculated from a weight measurement with analytical balance (KERN) for a known volume, taken with a pipette (Gilson).

\subsection{Verification of the system}

\begin{figure}
\resizebox{1\columnwidth}{!}{\includegraphics{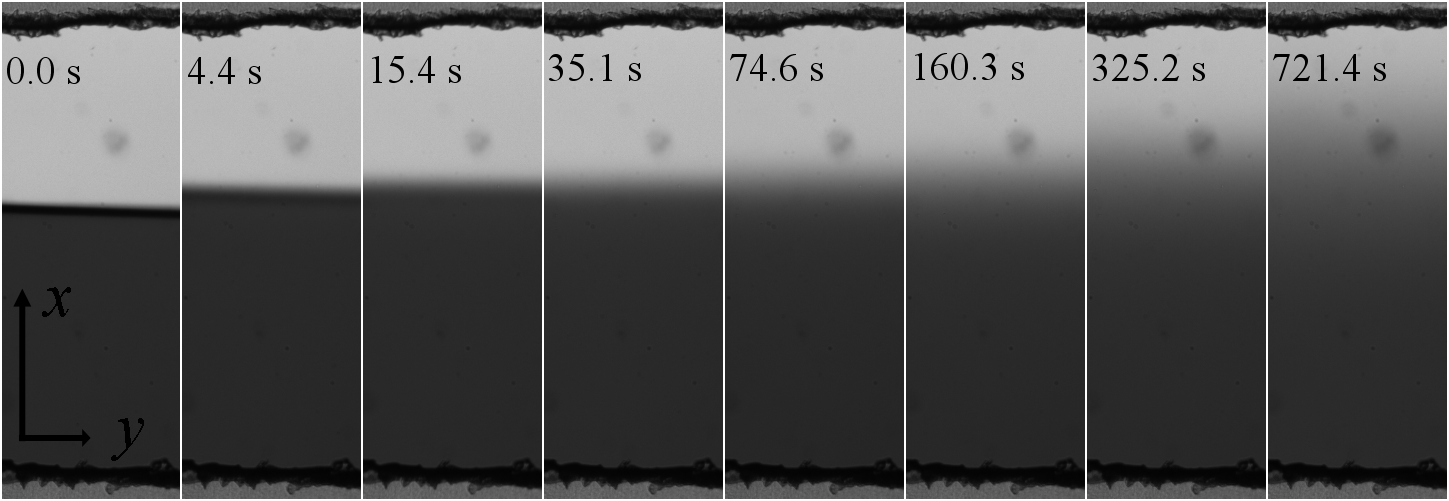}}
\caption{Snapshopts of magnetic particle diffusion across the microchannel with a stopped flow. Channel width $1.4$~mm.}
\label{fig:3}       
\end{figure}

As noted in the introduction, a parasitic convective motion even with no magnetic field present was influencing measurements in the previous experimental system.
It was due to the small density difference of the magnetic and non-magnetic fluids~\cite{JFM2,GKthesis}.
To eliminate it, the system has been redesigned and turned sideways, which allows the denser magnetic fluid to remain below water.
In order to verify it, a test measurement is performed, where only nanoparticle diffusion is expected.
For this, the flow of both fluids is suddenly stopped, obtaining a still system with a sharp initial interface.
The mixing process can be seen in the snapshots displayed in fig.~\ref{fig:3}.
A slow process is visible and the smearing gradually increases.

\begin{figure}
\resizebox{1\columnwidth}{!}{\includegraphics{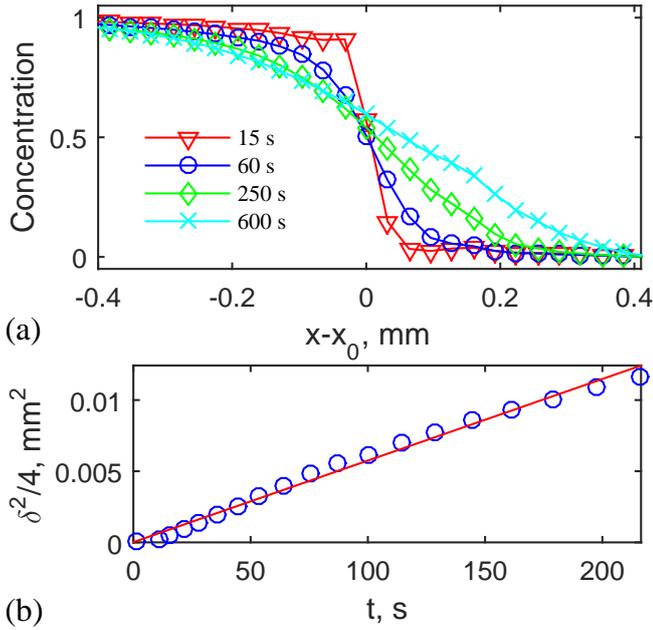}}
\caption{(color online) Analysis of nanoparticle diffusion test. Average concentration profiles at several time moments (15~s (red triangles), 60~s (blue circles), 250~s (green diamonds) and 600~s (cyan crosses)) are shown in (a). The magnetic particle diffusion coefficient is determined with a fit (red line) of the diffusion length data (blue circles), as shown in (b).}
\label{fig:4}       
\end{figure}

Recorded image series of the magnetic particle diffusion is analyzed for a manually selected area.
The images are converted to concentration plots via Lambert-Beer law and normalized to initial concentration $c_0=1$, as described previously in~\cite{JMMM}.
Then each image is averaged along the $y$-axis, obtaining the average concentration profile.
Several concentration profiles are shown in fig.\ref{fig:4}~(a).
Each average concentration profile $c(x)$ is fitted with the diffusion curve, according to Fick's law solution:
\begin{equation}
c(x)=\frac{1}{2}\left(1-\mathrm{erf}{\frac{\left(x-x_0\right)}{\delta}}\right),
\end{equation}
where $\mathrm{erf}$ is the error function and $x_0$ is the coordinate of the symmetry center and gives a degree of freedom for the fit. $\delta$ is the diffusion length, defined as
\begin{equation}
\delta=2\sqrt{Dt},
\label{eq:difflength}
\end{equation}
where $D$ is the diffusion coefficient and $t$ is the time diffusion is happening.
Eq.~\ref{eq:difflength} can be rewritten in the form $\delta^2/4=Dt$, which is more suitable for visualization and fitting, as Diffusion coefficient agrees with the slope.
That is done with experimental data, as shown in fig.\ref{fig:4}~(b).
Data follow a rather linear increase, which can be quantified, leading to a diffusion coefficient $D=5.7\cdot10^{-7}$~cm$^2$/s.
For comparison, an estimated value can be calculated by Stokes-Einstein equation:
\begin{equation}
D=\frac{kT}{3\pi\eta d},
\label{eq:StokesEin}
\end{equation}
where $k$ is the Boltzmann constant, $T=293$~K is the fluid temperature, $\eta=0.01$~P is the water viscosity and $d=7.0$~nm is the magnetic nanoparticle diameter, as given before.
The resulting $D=6.1\cdot10^{-7}$~cm$^2$/s agrees well with the experimentally determined value, verifying the experimental setup and allowing to proceed with the micro-convection experiments.

\subsection{Experimental observations}

We started the micro-convection experiments in the same manner as the previously described diffusion experiment.
First, the fluids are pumped through the chip and at some point the flow is stopped and magnetic field is turned on.
Unfortunately, the instability development gets distorted by microscopic flows in the $y$-axis direction, probably, due to small pressure differences in the channels or other effects arising from the application of magnetic field.
Without further improvements to the system, we were unable to fix this, therefore we proceeded with under flow-experiments.

\begin{figure*}
\resizebox{1\textwidth}{!}{\includegraphics{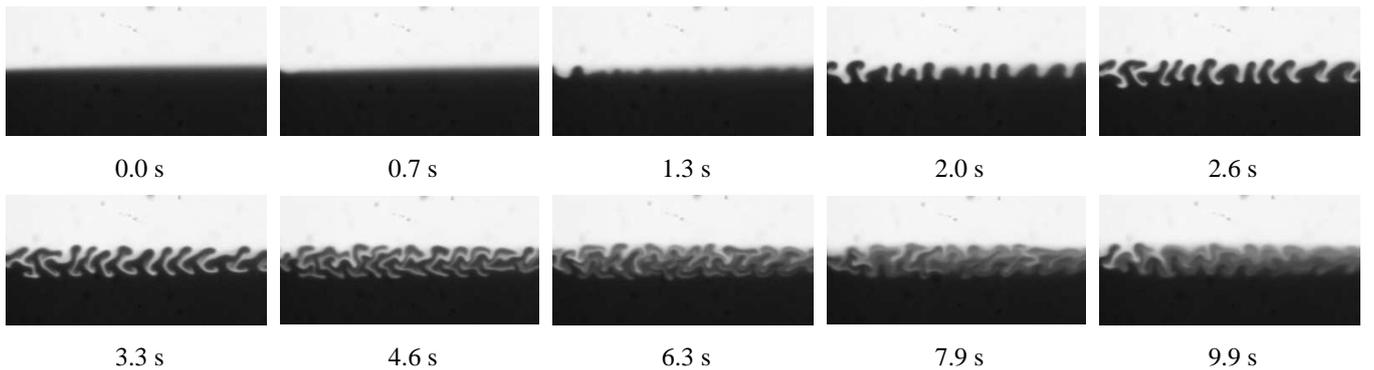}}
\caption{An example of magnetic micro-convection dynamics in a continuous microfluidics system. At $t=0.0$~s magnetic field is turned on. First, the instability is formed along the interface. Then, it is slowly smeared with the flow, going from left to right.  Finally, after $\approx10$~s a dynamic equilibrium state is reached, with an instability continuously forming on the fresh interface on the left side and quickly smearing on the way to the right side. Flow rate is $Q=1$ $\mu$l/min, magnetic field is $H=75$~Oe, volume fraction $\Phi_1=1.9\%$. Each image is $1.0\times2.0$~mm large.}
\label{fig:5}       
\end{figure*}

For each magnetic fluid we did experiments for different flow rates, selected on the syringe pump, and different magnetic fields, selected on the power supply.
An example of the dynamics registered is shown in fig.~\ref{fig:5} for a magnetic fluid with volume fraction $\Phi_1=1.9\%$ at a flow rate $Q=1$ $\mu$m/min and magnetic field $H=75$~Oe.
First, the flow rate is set and both fluids are let to flow.
Then the recording is started and the magnetic field is turned on ($t=0.0$~s in fig.~\ref{fig:5}).
At first, an initial instability with a development of clear and distinct fingers can be seen ($t=0.7...3.3$~s in fig.~\ref{fig:5}).
Then the flow distorts the growing finger pattern ($t=4.6...6.3$~s in fig.~\ref{fig:5}) and eventually reaches a situation that can be described as a dynamic equilibrium state ($t=7.9...9.9$~s in fig.~\ref{fig:5}), where new micro-convection instability forms at the tip of the Y-type channel, where magnetic and non-magnetic fluids meet and make a fresh interface (on the left side of images) and it continues to develop and becomes smeared, while being carried along the channel to the right side.
The process is recorded for 10-20 seconds, depending on the flowrate, so that a sufficient amount of data are recorded also for the dynamic equilibrium state.

After recording, magnetic field is then turned off and a sufficient time is waited for the fluids to form the initial no-field situation, before proceeding with the next field.

\begin{figure*}
\resizebox{1\textwidth}{!}{\includegraphics{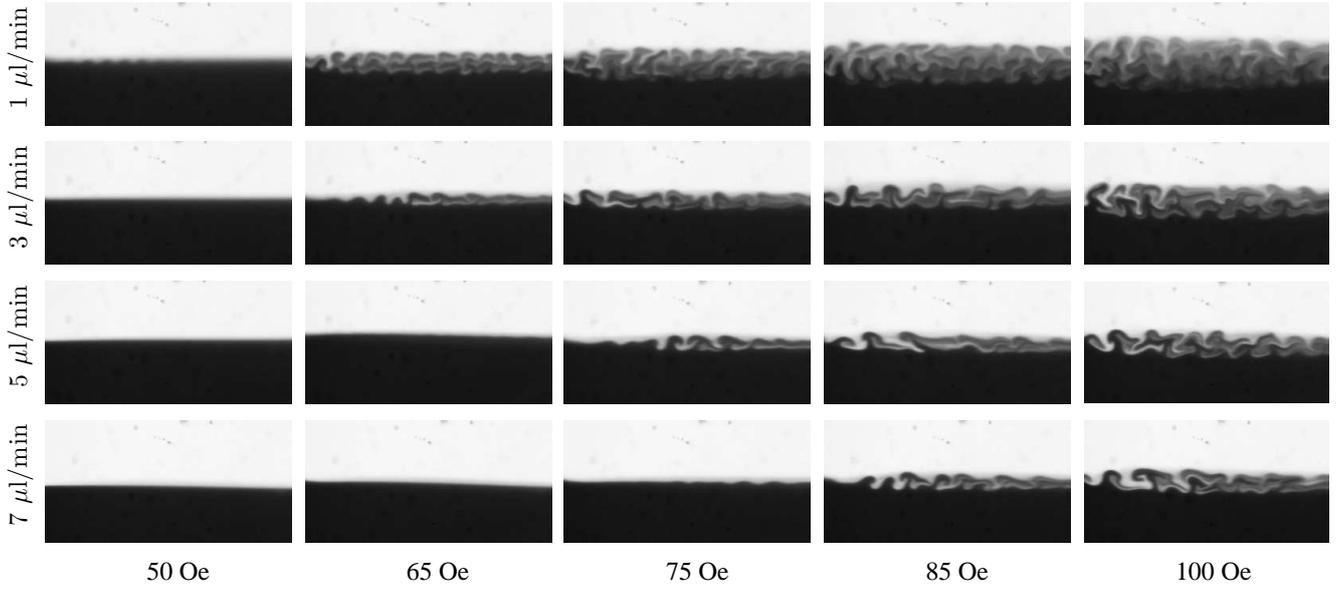}}
\caption{Dynamic equilibrium state images of magnetic micro-convection for various flow rates and magnetic fields. Magnetic fluid volume fraction $\Phi_1=1.9\%$. Each image is $1.0\times2.0$~mm large.}
\label{fig:6}       
\end{figure*}

Fig.~\ref{fig:6} gives an example of a dynamic equilibrium state images of magnetic micro-convection for a magnetic fluid with volume fraction ($\Phi=1.9\%$).
Each row represents a different flow rate, increasing from top to down, while each column has a different magnetic field, increasing from left to right.
One can observe that the magnetic micro-convection is noticeable only in a part of the images.
From previous works it is known that there is a critical field, below which the instability does not happen.
However, two conclusions can be made.
First, the critical field depends on the flow rate.
Second, the critical field in these conditions is $\approx10$ times larger than is expected by the current Brinkman model~\cite{JFM2}.

The first effect of the flow rate influence can be attributed to the laminar flow in the channel, that stabilizes the interface perturbations.
We can exclude it by gradually decreasing the flow rate and estimating the critical field at zero flow rate.
But the second effect must come from gravitational influence, which tries to keep the denser magnetic fluid below the less dense water and also stabilizes the interface.
And this effect we can not exclude, therefore, an improvement to our theoretical model is necessary.

The theoretical considerations are further developed in sect.~\ref{sec:3}.
They are based on a similar approach as was used for describing convective interface smearing in our previous paper \cite{JFM1}.

To provide comparable experimental data, we extend the experimental study to have four different dilutions of the original magnetic fluid: original volume fraction $\Phi_0=2.8\%$, slightly diluted $\Phi_1=1.9\%$, half diluted $\Phi_2=1.4\%$ and more diluted $\Phi_3=0.9\%$.
We do not go for smaller dilutions, as we approach both the limit of our electromagnet and the end of the linear magnetization regime for our magnetic fluid.
For each of the dilutions, we visually look for the tiny interface perturbations to detect the critical magnetic field.
This is repeated for every flow rate.
Diluting the magnetic fluid leads to a reduced density difference, but also to a reduced susceptibility.

\begin{figure*}
\resizebox{1\textwidth}{!}{\includegraphics{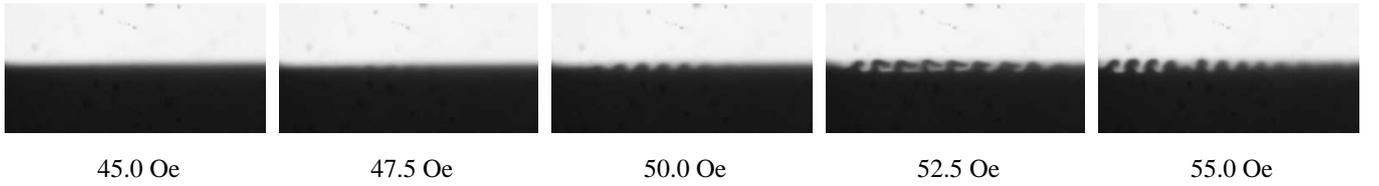}}
\caption{Visual detection of the critical magnetic field for a magnetic fluid with volume fraction $\Phi_1=1.9\%$ and a flow rate $Q=1$ $\mu$l/min. Critical field is $H=47.5$~Oe, as small interface perturbations can be seen.  Each image is $1.0\times2.0$~mm large.}
\label{fig:7}       
\end{figure*}

An example on the visual determination for magnetic fluid with volume fraction $\Phi_1=1.9\%$ in a flow rate $Q=1$ $\mu$l/mi is shown in fig.~\ref{fig:7}.
In this case the critical field is registered as $(H\pm\Delta H)=(47.5\pm1.3)$~Oe, as small interface perturbations can be seen, as compared to more pronounced perturbations for $H=50.0$~Oe or no perturbation for $H=45.0$~Oe.

\begin{figure}
\resizebox{1\columnwidth}{!}{\includegraphics{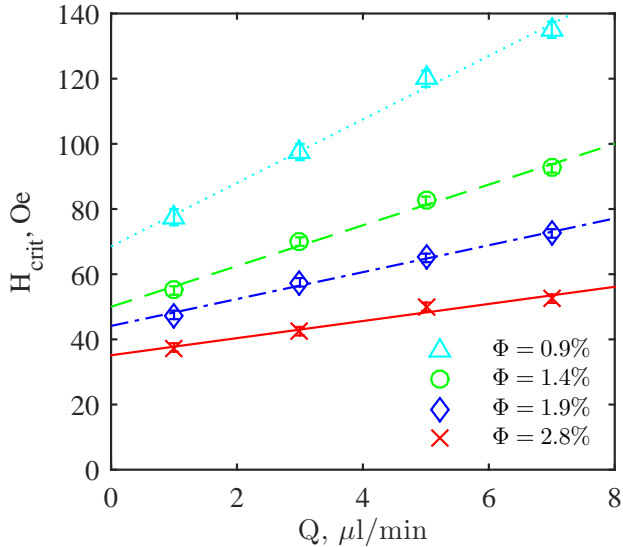}}
\caption{(color online) Critical magnetic fields for various flowrates and different dilutions of magnetic fluid. Lines correspond to linear fits that are used to extrapolate the critical magnetic field values at $Q=0$~$\mu$l/min.}
\label{fig:8}       
\end{figure}

The critical fields for the different dilutions of the magnetic fluid for different flow rates are summarized in the fig.~\ref{fig:8}.
Red crosses mark data points of the initial fluid ($\Phi_0=2.8\%$), blue diamonds mark slightly diluted magnetic fluid ($\Phi_1=1.9\%$), green circles are the half diluted fluid ($\Phi_2=1.4\%$) and cyan triangles correspond to the more dilute magnetic fluid ($\Phi_3=0.9\%$)
It seems that the critical field depends linearly on the flow rate.
But, as several handmade microfluidic channels have been used in experiments and with current method it is impossible to reproduce perfectly the channel shapes, this question is left for a following paper.
However, by fitting these curves and finding their intercept with vertical axis, using Least Squares method, we can find the expected critical magnetic fields and their errors for each of the fluids at zero flow.
For $\Phi_0=2.8\%$ it is $\left(H_{crit}\pm\Delta H_{crit}\right)_{0}=\left(35\pm2\right)$~Oe, for $\Phi_1=1.9\%$ it is $\left(H_{crit}\pm\Delta H_{crit}\right)_{1}=\left(44\pm1\right)$~Oe, for $\Phi_2=1.4\%$ it is $\left(H_{crit}\pm\Delta H_{crit}\right)_{2}=\left(50\pm2\right)$~Oe and for $\Phi_3=0.9\%$ it is $\left(H_{crit}\pm\Delta H_{crit}\right)_{3}=\left(69\pm2\right)$~Oe.
These values will be further used for comparison in sect.~\ref{sec:5}.

\begin{figure}
\resizebox{1\columnwidth}{!}{\includegraphics{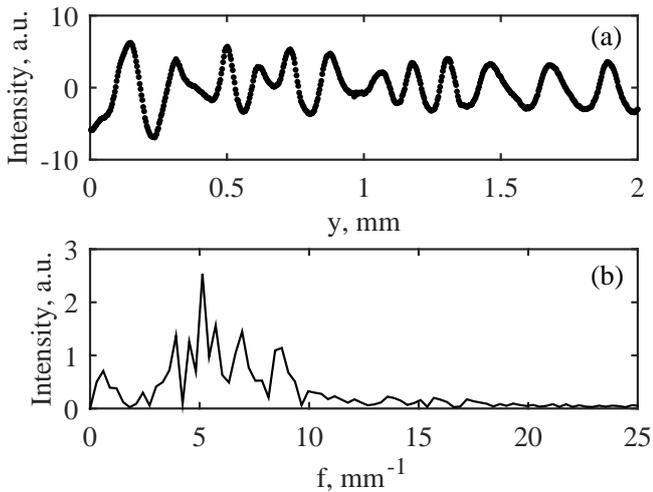}}
\caption{An example of image analysis (fig.~\ref{fig:5}, $t=2.0$~s) to determine the characteristic wavelength of the initial instability. (a) Average normalized $x$ intensity. (b) Spectrum of the average intensity. Peak is $5.2$~mm$^{-1}$.}
\label{fig:9}       
\end{figure}

We also investigate the characteristic size of the instability.
For that, the images of the initial instability are utilized, as characteristic fingers can be clearly distinguished.
They are recorded right after the magnetic field is turned on.
An example can be seen in fig.~\ref{fig:9}, where the image from fig.~\ref{fig:5} ($t=2.0$~s), is processed.
First, the average intensity $I_{\bar{x}}(y)$ is found by averaging along the $x$ axis.
Then this curve is normalized by subtracting a linear fit of the data in order to remove intensity bias in the original image. 
The normalized curve keeps the instability characteristics without additional biases, as can be seen in fig.~\ref{fig:9}~(a).
Afterwards, the finger frequency is obtained by applying a Fourier transform to the normalized curve.
Finding the peak in the resulting amplitude spectrum gives the characteristic frequency (see fig.~\ref{fig:9}~(b)), in this case $f_c=5.2$~mm$^{-1}$, which gives a characteristic wavelength $\lambda_c=0.19$~mm.

It is worth to note that the instability formation happens continuously at the initial contact point of the two fluids, which is at the tip of the Y channel junction, and can be called a at a dynamic equilibrium state.
These conditions are rather different from the theoretical model used later on, where instability develops at once across a long interface.
For these reasons we chose to find the characteristic size from the instability that develops across the flowing interface when the magnetic field is turned on, as shown in fig.~\ref{fig:5}.

\begin{figure}
\resizebox{1\columnwidth}{!}{\includegraphics{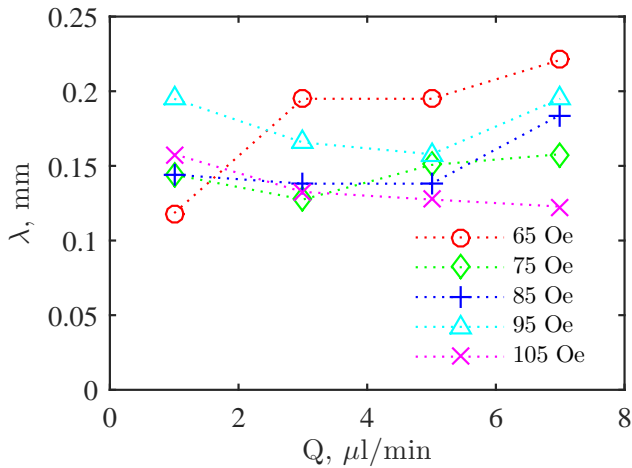}}
\caption{(color online) Characteristic wavelength of the initial instability at different magnetic fields measured for various flowrates. $\Phi_0=2.8\%$. Measurement error is larger than the differences measured ($\sigma=0.03$~mm).}
\label{fig:10}       
\end{figure}

We use this characterization method for all samples and find characteristic width $\lambda$ for all four fluid pairs at the various flow rates and magnetic fields, which are above the critical fields and have a sufficient fingering pattern.
It turns out that the wavelength of the initial instability is close to constant and has no clear dependence on the flow rate.
For example, as can be seen in fig.~\ref{fig:10}, for the water and magnetic fluid $\Phi_0=2.8\%$ pair for 5 different magnetic fields and 4 different flow rates has a characteristic width $\left(\lambda\pm\Delta\lambda\right)\approx\left(0.15\pm0.05\right)$~mm.
At least the measurement precision is insufficient to notice any dependence on the flow.

Therefore we calculate an average value $\bar{\lambda}$ for each magnetic field and use it as an estimate for the zero-flow ($Q=0$) case to compare it with theoretical results in section~\ref{sec:5}.

\section{Theoretical model}
\label{sec:3}

\subsection{Mathematical formulation}

\begin{figure}
\resizebox{0.85\columnwidth}{!}{\includegraphics{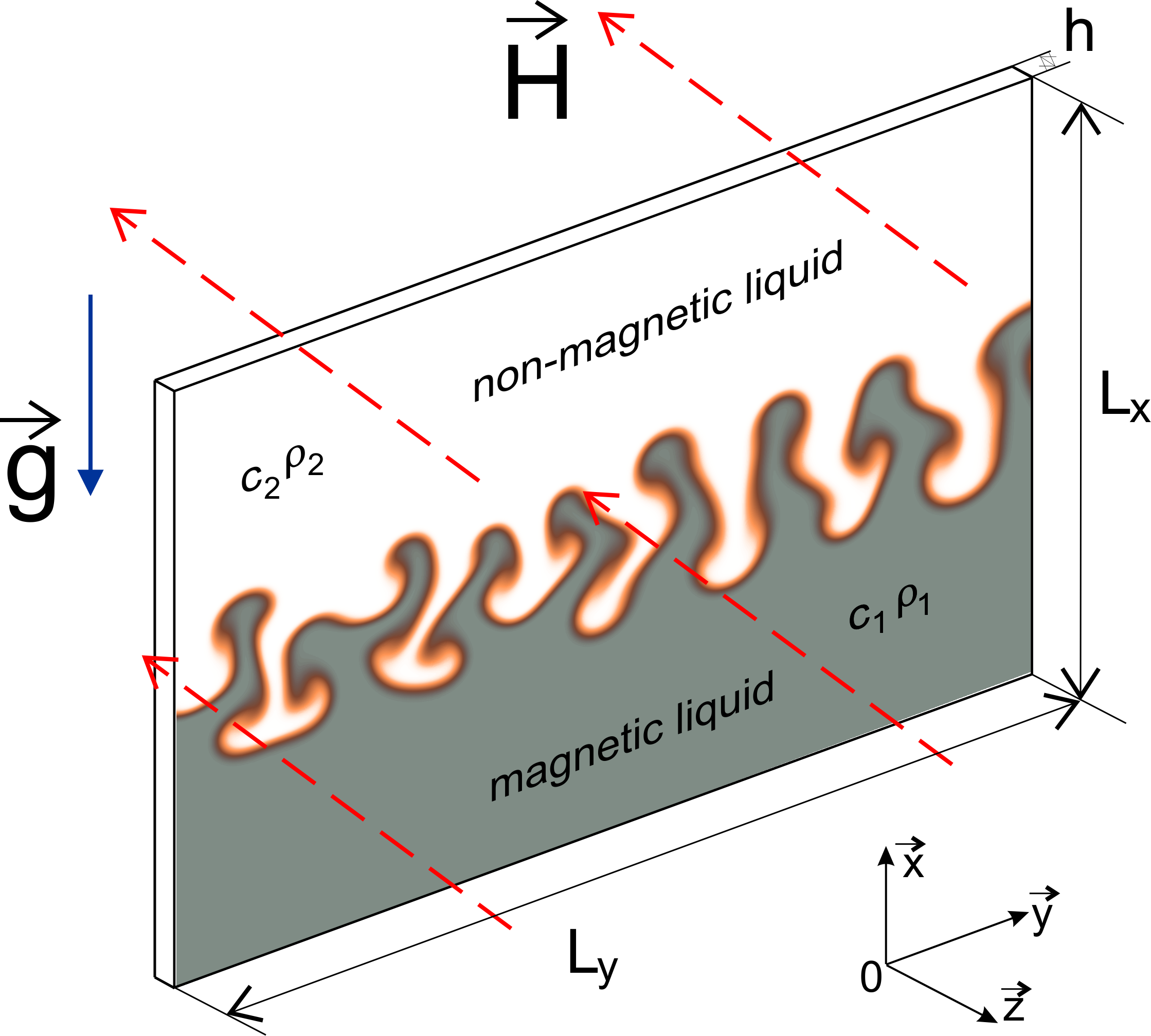}}
\caption{(color online) The sketch of a Hele-Shaw cell as considered in the model.}
\label{fig:heleshaw}       
\end{figure}

We consider two miscible fluids confined in a Hele-Shaw cell where the first fluid is a ferromagnetic fluid and the second is
a non-ferromagnetic fluid.
The Hele-Shaw cell is located vertically and the magnetic field is applied perpendicularly to the cell, as shown in the sketch in fig.~\ref{fig:heleshaw}.
The viscosities of the two fluids are considered equal.
Due to the ponderomotive forces of the non-homogeneous self-magnetic field on the interface between fluids the fingering instability arises.
The evolution of the fingering instability is described by a set of equations, which includes the Brinkman equation, the continuity and convection – diffusion equation \cite{Igonin,Cebers:97} and reads

\begin{gather}
\nonumber -\nabla p -\frac{12 \eta}{h^2}\boldsymbol{\rm v} - \frac{2 M(c)}{h}\nabla \psi_{\rm m}(c) + \eta {\rm\Delta}\boldsymbol{\rm v} + {\rm\Delta}\rho c \boldsymbol{\rm g}   = 0,\\
{\rm\nabla} \cdot \boldsymbol{\rm v} = 0,\\
\nonumber\frac{\partial c}{\partial t} + (\boldsymbol{v } \cdot \nabla) c = D \nabla^2 c~.
\label{Eg:Hele:2}
\end{gather}

\noindent where $\boldsymbol{v}=(v_{x},v_{y})$ is the depth averaged velocity, $p$ is pressure, $\eta$  is the viscosity of the fluid, $h$ is the thickness of  the Hele-Shaw cell,  $c$ is the concentration of magnetic fluid normalized by its value far from the interface, $D$ is the isotropic constant diffusion coefficient and ${\rm\Delta}\rho=\rho_1-\rho_2$ is the density difference between the fluids.
The magnetization $M(c)$ is taken to be proportional to the concentration of the magnetic fluid $c$ ($M = M_0 c$) and the value of the magnetostatic  potential  $\psi_{\rm m}$ on the boundary of the Hele-Shaw cell is given by \cite{Cebers:81,GoldstCebers:94}
$\psi_{\rm m}(\boldsymbol{r},t) = M_0\int c(\boldsymbol{r}^{'},t) K(\boldsymbol{r}-\boldsymbol{r}^{'},h) {\rm d} S^{'}$
where the integration is performed over the boundary of the Hele-Shaw cell, $ K(\boldsymbol{r},h)=1/\mid \boldsymbol{r}\mid-1/\sqrt{\mid\boldsymbol{r}\mid^2+h^2}$.

The boundary conditions for the velocity components and the concentration of the fluids  and the conditions of the periodicity across which require that the fluid is motionless at both ends of the Hele-Shaw cell are as follows:
\begin{gather}
 \label{Eg:Hele:Bound:1}
\nonumber v_{x}(0,y) = v_{y}(0,y)=0,~c(0,y)=1,\\
 v_{x}(L_{x},y) =v_{y}(L_{x},y)=0,~c(L_{x},y)=0~,\\ \nonumber
  \boldsymbol{v}(x,0,t) = \boldsymbol{v}(x,L_{y},t), \quad c(x,0,t) = c(x,L_{y},t)~.
\end{gather}

\indent The equations are put in dimensionless form by introducing the following scales:  length $h$, time $h^2/D$, velocity $D/h$, magnetostatic potential $M_{0}h$. As a result, the set of dimensionless equations reads

\begin{gather}
\nonumber  -\nabla p -  \boldsymbol{v} - 2 \textit{Ra}_{ m} c \nabla \psi_{\rm m} + \frac{\Delta \boldsymbol{v}}{12}  - \textit{Ra}_{g} c \boldsymbol{e}_x = 0,\\
\nabla \cdot\boldsymbol{v} = 0~,\\
\nonumber\frac{\partial c}{\partial t} + (\boldsymbol{v}\cdot\nabla) c =  \nabla^{2}c~.
    \label{Eg:HeleDim:3}
\end{gather}

\noindent Here $\textit{Ra}_{ m}$ is the magnetic Rayleigh number and  $\textit{Ra}_{g}$ is the gravitational Rayleigh number.
We have previously shown that $\textit{Ra}_{ m}$ is governing the magnetic micro-convection process \cite{JFM2}.
It is determined by the ratio of the characteristic time of the diffusion $\tau_{D}=h^{2}/D$ and the characteristic time of motion due to non-homogeneous self-magnetic field of the fluid $\tau_{M}=12\eta/M_{0}^{2}$, expressed as
\begin{equation}
\textit{Ra}_{ m}=M^{2}_{0}h^{2}/12\eta D~.
\label{eq:ram}
\end{equation}
The gravitational Rayleigh number is defined as  the ratio of the characteristic time of the diffusion $\tau_{D}=h^{2}/D$ and the characteristic time of motion due to  gravitational field  $\tau_{G}=12\eta/\Delta \rho g h$
\begin{equation}
\textit{Ra}_{g}=\Delta \rho g h^{3}/12\eta D~,
\label{eq:rag}
\end{equation}
where $\Delta\rho=\rho_1-\rho_2$ is the density difference between the denser fluid below and less dense fluid above and $g$ is the standard gravity.

\subsection{The linear stability analysis}

\begin{figure}
\resizebox{1\columnwidth}{!}{\includegraphics{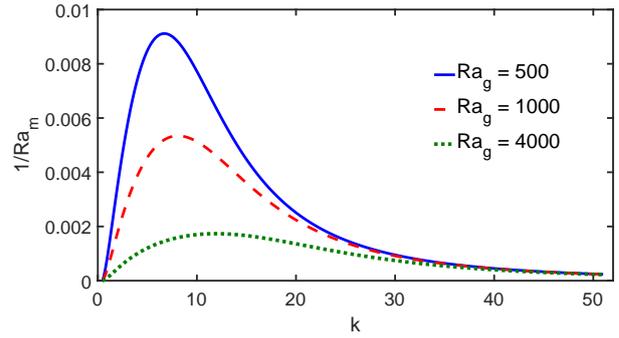}}
\caption{(color online) Neutral curves $\lambda = 0$ of magnetic micro-convection at a vertical Hele-Shaw cell for the Brinkman
model as obtained by the linear stability analysis at different values of the gravitational Rayleigh number.
Here $\textit{Ra}_{g} = 500$ (blue line), $\textit{Ra}_{g} = 1000$ (red dashed line), $\textit{Ra}_{g} = 4000$ (green dotted line).}
\label{fig:LinAnaliz::NeutralBm}       
\end{figure}

\begin{figure}
\resizebox{1\columnwidth}{!}{\includegraphics{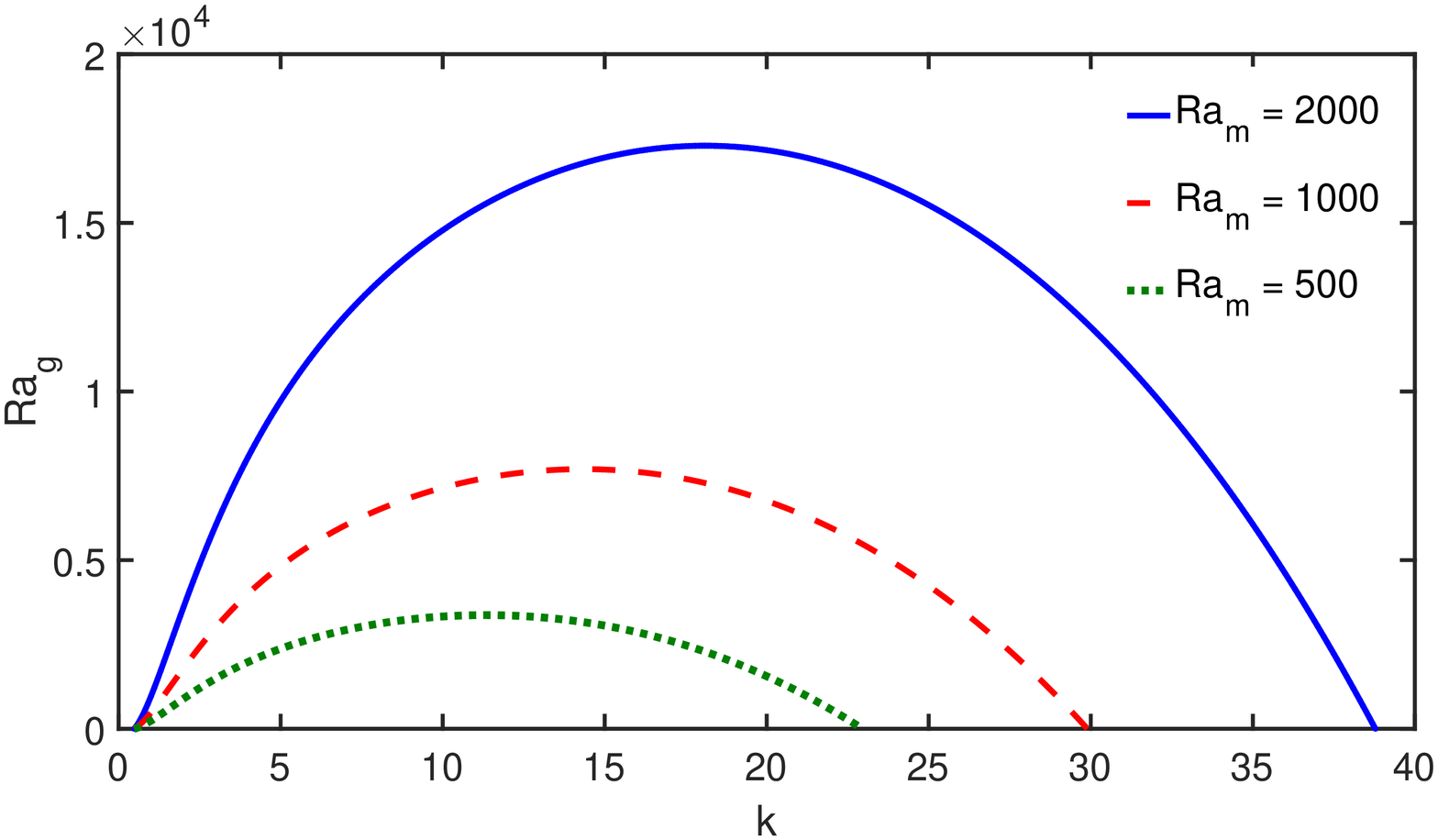}}
\caption{(color online) Neutral curves $\lambda = 0$ of magnetic micro-convection at a vertical Hele-Shaw cell for the Brinkman
model as obtained by linear stability analysis at different values of the magnetic Rayleigh number.
Here $\textit{Ra}_{m} = 500$ (green dotted line), $\textit{Ra}_{m} = 1000$ (red dashed line), $\textit{Ra}_{m} = 2000$ (blue line).}
\label{fig:LinAnaliz::NeutralBg}     
\end{figure}

\begin{figure}
\resizebox{1\columnwidth}{!}{\includegraphics{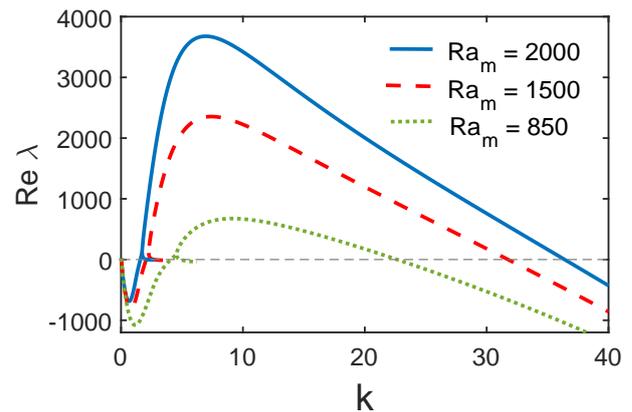}}
\caption{(color online) Growth increments of the instability at the interface of two miscible fluids as a function of the wavenumber for different magnetic Rayleigh numbers.
Here $\textit{Ra}_{m} = 2000$ (solid blue line), $\textit{Ra}_{m} = 1500$ (striped red line), $\textit{Ra}_{m} = 850$ (dotted green line) and gravitational Rayleigh number $\textit{Ra}_{m} = 4000$.}
\label{fig:LinAnaliz:Inkrement:1}     
\end{figure}

\begin{figure}
\resizebox{1\columnwidth}{!}{\includegraphics{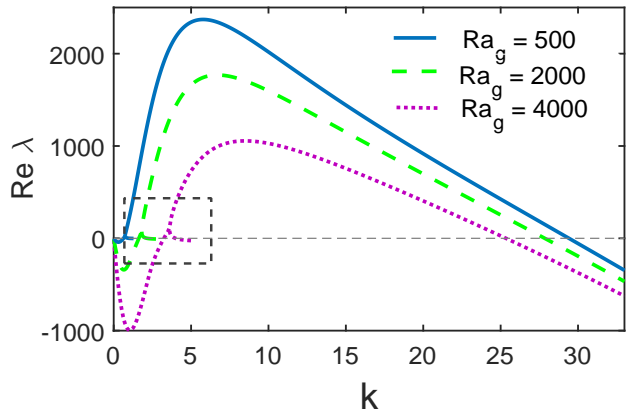}}
\caption{(color online) Growth increments of the instability at the interface of two miscible fluids as a function of the wavenumber for different gravitational Rayleigh numbers.
Here $\textit{Ra}_{g} = 4000$ (violet dotted line), $\textit{Ra}_{g} = 2000$ (stripped green line), $\textit{Ra}_{g} = 500$ (solid blue line) and magnetic Rayleigh number $\textit{Ra}_{m} = 1000$. Dashed area is shown in higher detail in fig.~\ref{fig:LinAnaliz:Inkrement:3}.}
\label{fig:LinAnaliz:Inkrement:2}     
\end{figure}

The linear stability analysis on the boundary of two fluids  is performed.  
An analytical solution may be found in the limit for smearing parameter $t_0 = 0 $ , that is, when the concentration distribution on the boundary between the two fluids is step-like. 
The quasi-stationary approximation for development of small perturbations is considered.
The linear perturbation of a quiescent base state is represented by  $\{c,\varphi_m,v_x,v_y\}(x,y,t) = \{c_0,\varphi_{m0},0,0\}(x) + \{c',\varphi_m^{'},v_x^{'},v_y^{'}\}(x)e^{iky + \lambda t}$. 
The dispersion relation reads

\begin{eqnarray}
\nonumber s k  +  \textit{Ra}_{ m}\Big(k\Big[\frac{g(k(s+m),\infty)}{m} - g(k(s+1),\infty)\Big]+
\\
+ 2\frac{m-1}{m}J(s,k) - \frac{\textit{Ra}_{ g}}{2}\Big[\frac{1}{m(s+m)}-\frac{1}{s+1} \Big]\Big) = 0
\label{Eg:LinAnaliz:Dispersion}
\end{eqnarray}

here  the parameters $s$ and $m$ are $ s = \sqrt{1+\lambda/k^2}$,  $m = \sqrt{1 + {12}/{k^2}} $  and the functions $J(p,q)$  and $g(a,z)$ are defined by the integrals
\begin{equation}
J(p,q)  = \int_{0}^{\infty} e^{-pz}(K_0(z)-K_0(\sqrt{z^2+q^2})){\rm d}z
 \end{equation}

\begin{equation}
 g(a,z)=  \int\limits_{0}^{z} e^{-a \zeta} \ln(1+\zeta^{-2}) {\rm d}\zeta
 \end{equation}

Here $K_0$ is the modified Bessel function of the second kind. The details of the solution in the limit $t_{0}\rightarrow 0$ is described in appendix A.

Solving the transcendental dispersion eq.~(\ref{Eg:LinAnaliz:Dispersion}) numerically at $\lambda = 0$ gives the neutral curves of the magnetic micro-convection for the Brinkman model.
For example, in fig.~\ref{fig:LinAnaliz::NeutralBm} the magnetic Rayleigh number is shown as a function of the wavenumber $k$ for different values of the gravitational Rayleigh number, $Ra_g = 500, 1000$ and $4000$.
In fig.~\ref{fig:LinAnaliz::NeutralBg}  the gravitational Rayleigh number is displayed as a function of the wavenumber $k$ for different values of the magnetic Rayleigh number, $Ra_m = 500, 1000$ and $2000$.
The critical values of magnetic and gravitational Rayleigh numbers are found at maxima of neutral curves in fig.~\ref{fig:LinAnaliz::NeutralBm} and fig.~\ref{fig:LinAnaliz::NeutralBg}.
An analysis of these curves show that the region above the maxima of the curves is unstable for developing micro-convection on the miscible interface. 
The instability can only be developed if the corresponding Rayleigh numbers are below the critical magnetic or gravitational Rayleigh numbers.
The corresponding dependence for $Ra_m$ and $Ra_g$ is used for comparison between the experimental and our theoretical model in fig.~\ref{fig:CompRamRag} in sect.~\ref{sec:5}.

Solving  eq.~(\ref{Eg:LinAnaliz:Dispersion}) gives the the growth increment as a function of the wavenumber $k$.
Fig.~\ref{fig:LinAnaliz:Inkrement:1} shows them for a fixed gravitational Rayleigh number $Ra_g = 4000$ with different values of the magnetic Rayleigh number,  $Ra_m = 500, 1000, 2000$, 
Fig.~\ref{fig:LinAnaliz:Inkrement:2} shows them for a fixed magnetic Rayleigh number $Ra_m = 1000$  and different gravitational Rayleigh numbers,  $Ra_g = 500, 2000, 4000$. 

At the initial time moment $t=0$ the mixing pattern between two miscible interfaces is determined by the fastest growing mode. 
The maximal growth increments for tested values of the magnetic Rayleigh numbers in fig.~\ref{fig:LinAnaliz:Inkrement:1} and fig.~\ref{fig:LinAnaliz:Inkrement:2} give us the approximate wavenumbers.
These are $k \approx 6 - 8 $ and depend on both magnetic and gravitational Rayleigh numbers $Ra_m$ and $Ra_g$.

On one side, increase of the magnetic field intensity in the vertical Hele-Shaw cell destabilizes the miscible interface and the intensity of finger growth increases. 
On the other side, increase of the role of the gravitational force decreases the intensity of finger growth.
From fig.~\ref{fig:LinAnaliz:Inkrement:2} we can see that indeed our model agrees that the gravitational force stabilizes the instability.

\begin{figure}
\resizebox{1\columnwidth}{!}{\includegraphics{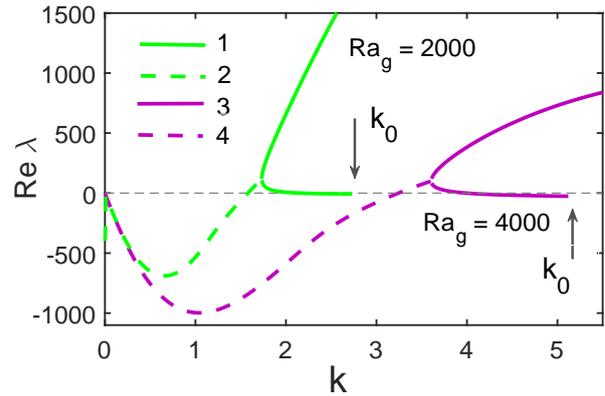}}
\caption{(color online) Magnification area of growth increments of the instability  in dependence on the wavenumber at two different values of the gravitational Rayleigh number $\textit{Ra}_{g} = 2000 (green), \textit{Ra}_{g} = 4000  (violet)$ and magnetic Rayleigh number $\textit{Ra}_{m} = 1000$. 1 and 3 (solid line) — the pair
of real increments; 2 and 4 (stripped line) — the real part of two complex-conjugate increments. }
\label{fig:LinAnaliz:Inkrement:3}     
\end{figure}

\begin{figure}
\resizebox{1\columnwidth}{!}{\includegraphics{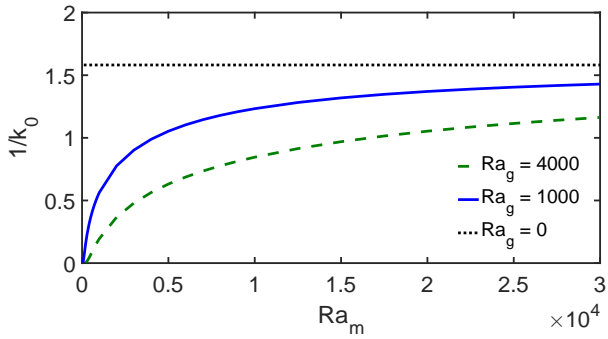}}
\caption{(color online) The wavenumber $k=k_0$ dependence on the magnetic Rayleigh number $\textit{Ra}_{m}$ at  different values of the gravitational Rayleigh number $\textit{Ra}_{g} = 0$  (black dotted line), $\textit{Ra}_{g} = 1000$ (blue line), $\textit{Ra}_{g} = 4000$ (green dashed line).}
\label{fig:LinAnaliz:Inkrement:4}     
\end{figure}

Numerically solving the dispersion relation eq.~(\ref{Eg:LinAnaliz:Dispersion}) in the case when the magnetic and gravitational  fields are applied may lead to a double-humped curve in the solution for growth increment, as can be better seen in fig.~\ref{fig:LinAnaliz:Inkrement:3}.
The presence of two preferred wavelengths may result in an interesting mode competition and interaction.
The solution of the transcendental dispersion eq.~(\ref{Eg:LinAnaliz:Dispersion}) for the growth increment as a function of the wavenumber $k$  show that there exists an area of the wavenumbers values where the growth increment is complex and can be in fig.~\ref{fig:LinAnaliz:Inkrement:3}.
That means that the instability of the interface between two miscible phases can have an oscillating character. 
The linear stability analysis is made  for a condition  ($\lambda \geq -k^2$) and for this condition the solution of the ordinary differential equation (ODE) is stable. 
For differential eq.~(\ref{Eq:LinAn:conc}) point where $k^2 = -\lambda$  gives a special point $k_0$ (indicated in fig.~\ref{fig:LinAnaliz:Inkrement:3}).
In these points the corresponding ODE solution does not satisfy the boundary conditions for perturbation of concentration at infinity and therefore will be unstable.
The dependence of the special point $k=k_0$ from the magnetic Rayleigh number $Ra_m$ is calculated and shown in fig.~\ref{fig:LinAnaliz:Inkrement:4}. 
In the limit $Ra_g \rightarrow 0$, the special point goes to $k_0 \rightarrow 0.62 $ and does not depend on the magnetic field intensity.

\section{Results and discussion}
\label{sec:5}
To compare experimental and theoretical results, we have to convert experimental values to dimensionless quantities.
For both magnetic Rayleigh and gravitational Rayleigh numbers, given by eqs.~(\ref{eq:ram}) and~(\ref{eq:rag}), we take channel thickness $h=0.013$~cm, viscosity $\eta=1$~P and diffusion coefficient as previously measured $D=5.7\cdot10^{-7}$~cm$^2$/s.
Additionally, for magnetic Rayleigh number $Ra_m$ we use magnetization $M_c=\chi\cdot H_c$, where $H_c$ is the critical field for zero flow determined in sect.~\ref{sec:2}, susceptibility $\chi$ can be calculated from the dilution factor $\chi_x=\frac{\chi_0\cdot\Phi_x}{\Phi_0}$ and $\Phi_0=2.8\%$ is the volume concentration of the original magnetic fluid.
And for gravitational Rayleigh number $Ra_g$ we need the density difference for fluid pairs.
This is measured with an analytical balance and a pipette, as described in sect.~\ref{sec:2}.
We find $\Delta\rho_0=0.148$~g/cm$^3$ for magnetic fluid with $\Phi_0=2.8\%$, $\Delta\rho_1=0.096$~g/cm$^3$ for $\Phi_1=1.9\%$, $\Delta\rho_2=0.071$~g/cm$^3$ for $\Phi_2=1.4\%$ and $\Delta\rho_3=0.045$~g/cm$^3$ for $\Phi_3=0.9\%$.
As the critical field values $H_c$ had a notable uncertainty, we also estimate the error, by taking $3\sigma$.

\begin{figure}
\resizebox{1\columnwidth}{!}{\includegraphics{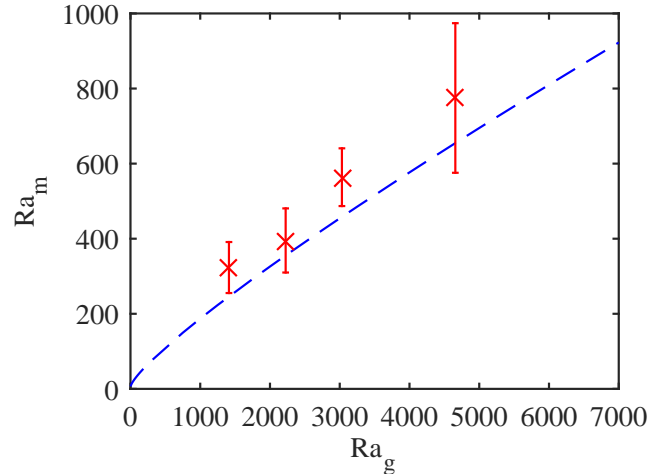}}
\caption{(color online) Comparison of critical Rayleigh numbers $Ra_g$ and $Ra_m$ between experimental and theoretical results. Data from linear analysis is given with a blue dashed line, while experimental points - red crosses and errorbars.}
\label{fig:CompRamRag}       
\end{figure}

From this, we can calculate that original magnetic fluid with $\Phi_0=2.8\%$ has $Ra_{g,0}=4657$ and $\left(Ra_m\pm Ra_m\right)_0=\left(775\pm199\right)$, the slightly diluted fluid with $\Phi_1=1.9\%$ has $Ra_{g,1}=3031$ and $\left(Ra_m\pm Ra_m\right)_1=\left(564\pm77\right)$, the half diluted fluid with $\Phi_2=1.4\%$ has $Ra_{g,2}=2225$ and $\left(Ra_m\pm Ra_m\right)_2=\left(395\pm85\right)$, but the more diluted fluid with $\Phi_3=0.9\%$ has $Ra_{g,3}=1412$ and $\left(Ra_m\pm Ra_m\right)_3=\left(323\pm68\right)$.
The experimental points are compared with pairs of critical $Ra_m$ and $Ra_g$ values obtained from linear stability analysis and are shown in fig.~\ref{fig:CompRamRag}. 
Data are in a reasonable agreement, confirming that the gravitational influence has a strong influence.

\begin{figure}
\resizebox{1\columnwidth}{!}{\includegraphics{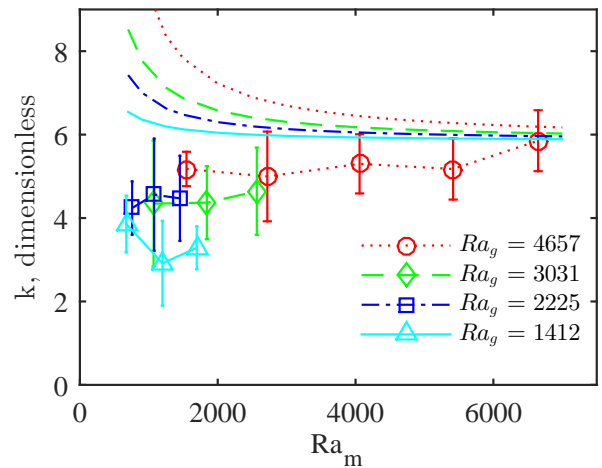}}
\caption{(color online) Comparison of characteristic wavenumber $k$ between experimental (open symbols and tracer lines) and theoretical results (only lines) as a function of magnetic Rayleigh number $Ra_m$ for several gravitational Rayleigh numbers $Ra_g$. Red dotted lines correspond to $Ra_{g}=4657$, green dashed lines - $Ra_{g}=3031$, blue dash-dotted lines - $Ra_{g}=2225$ and cyan lines - $Ra_{g}=1412$.}
\label{fig:WaveNumber}       
\end{figure}

We also compare the characteristic wavenumbers $k$ for the instability, as is shown in fig.~\ref{fig:WaveNumber}.
Theoretical results are obtained from the growth increment data, where the fastest growing mode is chosen, while experimental data are averaged for each fluid pair over different flow rates, as described earlier, and over similar magnetic field values, to improve the quality of the results.
In experiments with diluted fluids the range of accessible $Ra_m$ values is limited due to technical limitations of the coil system, susceptibility and its magnetization linear regime.
The values are of the same order of magnitude, characteristics and sequence of the data are similar, showing that sizes are close to constant values. 
However, differences in the absolute values and trends can be noted and need a further investigation.
It is worth to mention that all results in our theoretical model are given in the limit of the smearing parameter $t_0 = 0 $.
Studies of a similar system of magnetic microconvection in Hele-Shaw cell by Brinkmann  approximation have given analytical results for $t_0 > 0 $ of the smearing parameter ~\cite{Igonin}, showing that the wavenumber $k$ decreases if the initial smearing parameter $t_0$  was increased.
It is exactly what we see with our data and might be among the main reasons for the notable differences.
Of course, an improved theoretical model or experimental data at a zero-flow condition would help in explaining them. 

The difference in the experimental system is that the smearing time should be different along the interface, as during the time after Y junction in the microfluidics channel the diffusion takes place.
Hence, we can use the residence time in the field of view of the channel as an estimation of the smearing time.
For the slowest flow rate $Q=1$~$\mu$l/min, the metric velocity in $y$ direction is $v_y\approx0.13$~mm/s and to cross the $\Delta y=2.0$~mm field of view $\tilde{t}_{0,max}=\Delta y/v_y=15$~s.
In dimensionless units it is $t_{0,max}=\tilde{t}_{0,max}\cdot D/h^2=0.05$ and the smearing time ranges from $0$ to this value.
These are small smearing times $t_0$ and are similar to previous experiments \cite{JFM2,GKthesis}. 
We plan to investigate this in a higher detail in a future study that will include both numerical simulations and experiments for various $t_0$ values.

It is important to highlight that the gravitational Ray-leigh numbers in this situation are large.
Here it comes from a combination of medium sized microfluidic channels and magnetic colloids.
As formula shows, the gravitation effects can be almost excluded, if the thickness of the cell is 10 times smaller, because the thickness has a cube dependence, resulting in a reduction of the $Ra_g$ by $1000$ times.
In addition, typical magnetic colloids have a notable $5..15\%$ density difference with the carrying liquid while particles have a large enough diffusion coefficient not to sediment that quickly.
Although magnetic colloids are quite specific field, these conditions will be important both in the perspective industrial applications of magnetic fluids in thermoelectricity~\cite{ThermoDiff}, as well as for high throughput microfluidics systems in biotechnology.
Therefore, whenever similar conditions are met, the scientists must take gravitational aspect into account.

To estimate importance of a convective motion, one usually calculates the P\'{e}clet number $Pe=L\cdot u/D$, where $L$ is a characteristic scale, $u$ is convective velocity and $D$ is the diffusion coefficient.
In our case velocity due to gravity can be estimated as $u=h/\tau_G$, where $\tau_G=12\eta/(\Delta\rho g h)$ is the characteristic time of motion due to gravitational field and $h$ is the thickness of the cell.
If we take $L=h$ and put this it in the equation for P\'{e}clet number, we get
\begin{equation}
Pe=\frac{\Delta\rho g h^3}{12\eta D}=Ra_g
\end{equation}
that shows that P\'{e}clet number and gravitational Rayleigh number can be considered the same in our system.
If a different characteristic scale is taken, for example, width of the channel $w$, then the link between both dimensionless quantities is a scale factor $Pe=Ra_g\cdot w/h$.

Finally, the magnetic micro-convection is an effective and simple active mixer, as can be seen in the obtained images and has been justified by previous results~\cite{JMMM}.
However, a better quantification and predictability in conditions where gravity plays an important role is needed, what will be continued in a future study.

\section{Conclusions}
\label{sec:6}
In this study we have shown how gravitational effects can be important even for microscopic systems with small density differences.
First, we have experimentally demonstrated how gravity and fluid flow stabilizes the instabilities formed by magnetic micro-convection in a vertical channel.
Second, we have improved the theoretical model of magnetic micro-convection, based on the Brinkman equation, to describe the system at a zero-flow condition.
Third, we have used the critical field and characteristic width to compare the experimental and theoretical results, which show a reasonable agreement.

These results provide a basis for further research on various stability conditions, efficiency of the magnetic micro-mixer and theoretical concepts connecting them, as well as their applications.

\section*{Acknowledgments}
G. Kitenbergs' research has been funded by a project from PostDocLatvia Nr.1.1.1.2/VIAA/1/16/197.

We are grateful to the action MP1305 "Flowing matter", supported by COST (European Cooperation in Science and Technology).

Additionally, we would like to acknowledge the French-Latvian bilateral program "Osmose" (FluMaMi project) for sustaining collaboration with PHENIX laboratory. 
In particular, we thank R. Perzynski and E. Dubois for fruitful discussions and D. Talbot for preparing the magnetic fluid.

We thank M.M. Maiorov from Institute of Physics of the University of Latvia for magnetization measurements.

\section*{Appendix A}

Taking into account what in the limit $t_{0}\rightarrow 0$ the function  $\partial \psi_{\rm m0}(x,0)/\partial x   =  -\ln(1+x^{-2}) $ and as result the equations for the velocity perturbations
\begin{eqnarray}
\label{Eq:LinAn:veloc}
\left( \frac{\partial^2}{\partial x^2} -  k^2 \right )^2  v_x^{'} -{12}\left( \frac{\partial^2}{\partial x^2} -  k^2 \right )  v_x^{'}  - \\ \nonumber
-  24  k^2 \textit{Ra}_{ m} \ln(1+x^{-2})c^{'}  - 12 k^2 \textit{Ra}_{g} c^{'} = 0
\end{eqnarray}

and for concentration perturbations reads

\begin{eqnarray}
\label{Eq:LinAn:conc}
 (\lambda + k^2)c^{'} + v_x^{'} \frac{\partial c_0}{\partial x} - \frac{\partial^2 c^{'}}{\partial x^2}  = 0 ~.
\end{eqnarray}

The boundary conditions at the discontinuity of the concentration $c_{0}$ are given by continuity of the concentration perturbation, tangential and normal to the front velocity components and their derivatives:

\begin{eqnarray}
\label{Eq:LinAn:BC}
c^{'}(0^{+}) - c^{'}(0^{-}) = 0~,
\\ \nonumber
v^{'}_{x}(0^{+}) - v^{'}_{x}(0^{-}) = 0~,
\\ \nonumber
\frac{{\rm d}v^{'}_{x}}{{\rm d}x}(0^{+}) - \frac{{\rm d}v^{'}_{x}}{{\rm d}z}(0^{-}) = 0~,
\\ \nonumber
\frac{{\rm d}^{2}v^{'}_{x}}{{\rm d}x^{2}}(0^{+}) - \frac{{\rm d}^{2}v^{'}_{x}}{{\rm d}x^{2}}(0^{-}) = 0~,
\end{eqnarray}

The two lacking boundary conditions at the discontinuity are obtained by integration $\int_{-\delta}^{\delta} (\ldots) {\rm d}x $ of eq.s~(\ref{Eq:LinAn:veloc})-(\ref{Eq:LinAn:conc}) across the diffusion layer and taking the limit $\delta \rightarrow 0$ which gives
\begin{eqnarray}
\label{Eq:LinAn:BC1}
\frac{{\rm d}c^{'}}{{\rm d}x}(0^{+})-\frac{{\rm d}c^{'}}{{\rm d}x}(0^{-})=-c_0 u_{x}^{'}(0)~,
\\ \nonumber
\frac{{\rm d}^{3}u^{'}_{x}}{{\rm d}x^{3}}(0^{+})-\frac{{\rm d}^{3}u^{'}_{x}}{{\rm d}x^{3}}(0^{-}) = -24  k^2 {Ra}_{\rm m} \psi_{\rm m}^{'}(0) ~.
\end{eqnarray}
From  eq.~(\ref{Eq:LinAn:conc}) and boundary condition eq.~(\ref{Eq:LinAn:BC})  for concentration $c^{'}$  and taking into account condition at infinity $c^{'}(\infty) = 0, c^{'}(-\infty) = 0$   follows that $c^{'} |_{x<0} =  Q \displaystyle e^{\displaystyle \sqrt{\lambda + k^2} x }$ and $ c^{'} |_{x>0} =  Q \displaystyle e^{\displaystyle -\sqrt{\lambda + k^2} x }$. The general  solution  of the Eqs.~(\ref{Eq:LinAn:veloc}) reads
\begin{eqnarray*}
\label{Eq:LinAn:vel:homsol}
\qquad  {u}^{'}_{x} = \tilde{A}_1(x) e^{ k x } + \tilde{B}_1(x) e^{ -k x } +  \\ \nonumber
+\tilde{C}_1(x) e^{\sqrt{k^2 + 12} x } + \tilde{D}_1(x) e^{-\sqrt{k^2 + 12} x }~.
\end{eqnarray*}
where the functions  $\tilde{A}_1, \tilde{B}_1, \tilde{C}_1, \tilde{D}_1$  are given by the solution of the set of linear differential equations and a result the solution for velocity reads
\begin{eqnarray*}
{v}^{'}_{x}|_{x<0} = {A_1} e^{ k x } + {B_1} e^{ -k x } +{C_1} e^{k m x } + {D_1} e^{-k m x }
 \\
-{k}{} Q w(-k(s-1),x)e^{ k x } +{k}{}  Q w(-k(s+1),x)e^{ -k x }+
 \\
+\frac{k}{m} Q w(-k(s-m),x)e^{k m x } - \frac{k}{m}  Q w(-k(s+m),x)e^{-k m x }
\end{eqnarray*}

\begin{eqnarray*}
{v}^{'}_{x}|_{x>0} = {A_2} e^{ k x } + {B_2} e^{ -k x } + {C_2} e^{k m x } + {D_2} e^{-k m x } +
 \\
-{k}{} Q w(k(s+1),x) e^{ k x } +{k}{}  Q w(k(s-1),x),x)e^{ -k x }+
 \\
+\frac{k}{m}  Q w(k(s+m),x)e^{k m x } - \frac{k}{m} Q w(k(s-m),x)e^{-k m x }
\end{eqnarray*}
where
\begin{eqnarray*}
 w(a,z)=  Ra_m g(a,z) + \frac{Ra_g}{2}f(a,z)
\end{eqnarray*}

Boundary conditions eq.(\ref{Eq:LinAn:BC}) and the conditions of vanishing perturbation at infinity gives the set of linear algebraic equations, which condition of solubility gives the dispersion equation for the growth increment of perturbations eq.(\ref{Eg:LinAnaliz:Dispersion}).

%
%
\section*{Authors contributions}
G.K. observed the phenomenon and designed the experimental system. G.K. and L.P. performed the experiments and analyzed the data. A.T. updated the theoretical model and performed the stability analysis. A.C. formulated the model and supervised the study. G.K. and A.T. wrote the manuscript. All the authors were involved in improving the manuscript.
All the authors have read and approved the final manuscript.
%
%

\end{document}